\documentclass[twocolumn,pre]{revtex4}

\usepackage{graphics}
\usepackage{graphicx}
\usepackage{amsfonts}
\usepackage{textcomp}
\usepackage{amssymb}
\usepackage{mathrsfs}
\usepackage{amsmath}
\usepackage{color}
\usepackage{ulem}

\begin{document}

\title{Self-organization of active particles on random networks}

\author{Hanqing Nan$^1$, Yu Zheng$^2$, Shaohua Chen$^3$, Christopher Z. Eddy $^{4}$, Jianxiang Tian$^{5, 1}$, Wenxiang Xu$^{6, 1,*}$, Bo Sun$^{4, *}$ and Yang Jiao$^{1, 2,}$}
\email{sunb@onid.orst.edu (B. S.); xwxfat@gmail.com (W. X.);
yang.jiao.2@asu.edu (Y. J.)} \affiliation{$^1$Materials Science
and Engineering, Arizona State University, Tempe, AZ 85287;
$^2$Department of Physics, Arizona State University, Tempe, AZ
85287; $^3$Department of Materials Engineering, KU Leuven,
Kasteelpark Arenberg 44 Bus 2450, Leuven, Belgium; $^4$Department
of Physics, Oregon State University, Corvallis, OR 97331;
$^5$Department of Physics, Qufu Normal University, Qufu 273165, P.
R. China; $^6$College of Mechanics and Materials, Hohai
University, Nanjing 211100, P.R. China}

\begin{abstract}
%3D migration in collagen is important

%cell generate force, and also sense the mechanical cues via durotaxis.

%However the role of force network on the collective behavior is not clear.

%Here, we investigate collective migratory behaviors of active particles regulated by
%dynamically evolving force networks via local durotaxis using a minimal model.

%Present the model

%dynamic phase transition,

%quantitively consistent with experimental

%underlying physics?

Collective cell migration in 3D extracellular matrix (ECM) is
crucial to many physiological and pathological processes.
Migrating cells can generate active pulling forces via actin
filament contraction, which are transmitted to the ECM fibers and
lead to a dynamically evolving force network in the system. Here,
we elucidate the role of such force network in regulating
collective cell behaviors using a minimal
active-particle-on-network (APN) model, in which active particles
can pull the fibers and hop between neighboring nodes of the
network following local durotaxis. Our model reveals a dynamic
transition as the particle number density approaches a critical
value, from an ``absorbing'' state containing isolated stationary
small particle clusters, to a ``active'' state containing a single
large cluster undergone constant dynamic reorganization. This
reorganization is dominated by a subset of highly dynamic
``radical'' particles in the cluster, whose number also exhibits a
transition at the same critical density. The transition is
underlaid by the percolation of ``influence spheres'' due to the
particle pulling forces. Our results suggest a robust mechanism
based on ECM-mediated mechanical coupling for collective cell
behaviors in 3D ECM.

%The predicted phase transition is subsequently verified using {\it
%in vitro} metastatic breast cancer systems

\end{abstract}

%g_2 based translational order metric

%\pacs{05.20.-y, 61.43.-j}
%may need to

\maketitle

%\section{Introduction}

%talk about the complex process of 3D cell migration,

%mention individual cell migration rules, e.g., durotaxis, Studies
%suggest cell migration can be affected by micro-environment, but
%not put together, until very recently

%for ECM-cell migration mechanisms, cite both experimental work and simulations

%{\bf Add section titles in the print, make the organization very
%lear}

%{\bf it might be over complicated to mention the periodicity of
%the clusters, just use the occupation probability should be
%sufficient for illustrating the dynamics, this can be complemented
%by using the fraction of dynamic particles.}

%completely remove occupation probability, use

Collective cell migration is crucial to many physiological and
pathological processes such as tissue regeneration, immune
response and cancer progression \cite{ref1, ref2, ref3, ref4}.
Cell migration in 3D extracellular matrix (ECM) is a complex
dynamic process involving a series of intra-cellular and
extra-cellular activities \cite{ref12, ref13}, and can be
regulated by a variety of cell-ECM interactions via chemotaxis
\cite{ref14}, durotaxis \cite{ref16}, haptotaxis \cite{ref17}, and
contact guidance \cite{ref18}. A migrating cell also generates
active pulling forces, which are transmitted to the ECM fibers via
focal adhesion complexes \cite{ref21} and consistently remodel the
local ECM (e.g., by re-orienting the collagen fibers, forming
fiber bundles and increasing the local stiffness of ECM)
\cite{ref24, ref29, chen2019}. In a multi-cell system, the pulling
forces generated by individual cells can give rise to a
dynamically evolving force network (carried by the ECM fibers)
\cite{ref8, Frey07}. The force network can further influence the
migration of the cells, which in turn alters the ECM and the force
network \cite{ref5, ref6, ref7, ref8, ref9, ref10, ref11}. This
feedback loop between the force network and cell migration could
lead to a rich spectrum of collective migratory behaviors.

%Recent studies have indicated that a delicate balance among the
%magnitude of the pulling forces, the cell-ECM adhesion strength,
%and ECM rigidity is required to achieve an optimal mode of single
%cell migration \cite{ref30}.

The cell-ECM system is an example of complex many-body systems in
which the individuals (e.g., migrating cells) communicate and
interact with one another through their environment (e.g., ECM),
while simultaneously re-shaping the environment, altering the
means (e.g., the force network) to pass information among
themselves. Other examples of such complex systems include flocks
of birds, schools of fish, and active swimmers in crowded
environment \cite{Popkin16, Bechinger16}. In this letter, we
investigate the collective cellular dynamics and self-organizing
multi-cellular patterns in 3D ECM resulted from the dynamically
evolving force network, using a minimal active-particle-on-network
(APN) model. Although focusing on the cell-ECM system, the
physical insights obtained here are also valuable to understanding
active-particle systems dominated by environment-mediated
particle-particle interactions.

%{\bf The model can be adapted and insights valuable for other
%systems.}

%Although guided cell migration via ECM-mediated mechanical cues
%has recently been investigated \cite{ref31}, the evolution of the
%force network in ECM due to multi-cell migration, its role in
%regulating collective migratory behaviors, and in the emergence of
%possible self-organizing multi-cellular patterns remain to be
%elucidated.

%%%% discussion for general physics
%{\bf The system represents an example in which individuals
%communite with one antoehr via environment, and constantly
%re-shape the envinvorment (cite review). It is important to
%ecludieate the role}

%Indeed, in a recent {\it in vitro} experiment, an abnormal and
%rapid aggregation of invasive breast cancer cells (in contrast to
%random spreading) during a massive polarized invasion has been
%observed, which was resulted from the strong ECM-mediated
%mechanical coupling among the cancer cells induced by the
%collective polarization of the cells [*** our PRL].

%how the dynamic force network regulates

%what types collective migratory behaviors can result from force
%network

%experimental evidence of ECM remodeling and ECM mediated
%collectiveness

\begin{figure}[ht]
\includegraphics[width=0.4\textwidth,keepaspectratio]{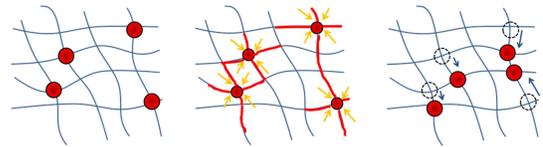}
\label{fig_1}\caption{Schematic illustration of the
active-particle-on-network (APN) model. Left: Particles on a
stress-free network. Middel: Particle contraction leads to a
``force network'' composed of high-stress fibers (illustrated
using red). Right: Particles migration on the ECM network along
fibers carrying the largest forces.}
\end{figure}

%A realization of the random network model generated based on
%maximally random jammed sphere packing. (c) Force network (carried
%by the high-stress fibers) generated by contractile particles
%(shown as red spheres) on the network. For better visualization,
%only a small sub-network is shown here.

{\bf Active-particle-on-network model:} The 3D ECM is modeled as a
discrete network with a ``graph'' (i.e., node-bond) representation
in a cubic simulation domain with linear size $L$, which is
composed of $M_n$ nodes and $M_b$ bonds. The average coordination
number $Z$, i.e., the average number of bonds connected to each
node, is given by $Z = 2M_b/M_n$. We have used both the periodic
boundary (PB) conditions and fixed boundary (FB) conditions (i.e.,
the nodes within a certain distance $\delta L$ from the boundaries
of the simulation domain are fixed) in our simulations, and find
that even for a moderate system size (e.g., $M_n \sim 5000$) the
boundary conditions do not affect the results. In the subsequent
discussions, we will mainly present the results obtained using the
fixed boundary conditions, under which $M_n$ denotes the number of
free (non-fixed) nodes.

%Talk about the boundary conditions, fixed and/or periodic, point
%out for large ones, the bc does not affect the results. point out
%Mn is free node for fixed bc, which will be used for the rest of
%the work.

%Then mention the active particles, contraction delta, causing a
%force

%active particles can occupy the nodes of a model network, pull the
%fibers connected to the occupied node, and hop between neighboring
%nodes following local durotaxis.

Next, $N_p$ active particles (e.g., congruent spheres) are
introduced in the network such that each particle occupies a
randomly selected un-occupied node (i.e., each node can be
occupied by only one particle). The {\it number density} $\rho$ of
the particles is defined as $\rho = N_p/M_n$, i.e., the fraction
of nodes occupied by the active particles. Each particle can
generate a contraction, which pulls all of the bonds connected to
the node it occupies towards the particle center (i.e., the node)
by a fixed amount $\delta l$, leading to different pulling forces
in the bonds and thus, a force network in the system. We consider
the particles can ``migrate'' from its original node to an
un-occupied neighboring node following local durotaxis, i.e.,
along the bond with the highest stiffness, which is also the bond
that carries the largest pulling force among all neighboring bonds
(see Fig. 1 for illustration). The diameter of the particles is
not essential in our model and thus, is not explicitly considered
here.

The bonds of the network are modeled as elastic elements with only
non-zero stretching modulus $E_s$ and free to rotate at the nodes.
An active particle can generate pulling forces in the bonds
connected to the node it occupies by contraction, i.e., $\delta
l$. This contraction leads to a strain $\epsilon_i = \delta l/l_i$
in the bond $i$ with original length $l_i$, and thus, a pulling
force $f_s = E_sA\delta l/l_i$, where $A$ is an effective
cross-sectional area of the bonds. These pulling forces impose
force boundary conditions for the ECM network, and the
force-balance network configuration is obtained using an iterative
force-based relaxation approach \cite{ref11}.

%but should be smaller than the shortest bond in the network.

%{\it Fig. 1: 3 steps of the hopping rule: original configuration;
%contraction that deform the network, and generate high stress
%bonds; particle hopping event following local durotaxis. Then show
%a pure network picture, a network with randomly distributed
%particles; and lastly a force network in this system.}

%Here, we employ a linear elastic network, for the following
%reasons: XXX.

%Mention in the Method section the solution method for obtain mechanical equilibrium
%network configurations, using conjugate gradient method.

%point out both cell distribution and network topology/geometry
%will affect the results, and we have explored both.

%More specifically, both of these factors can affect the
%organization of the force network, also the properties of the
%network fibers, cell contractibility etc. Here, we make a clear
%model, assuming linear elasticity of the network fibers, so
%everything is superposition. Once the fundamentals are clear, the
%other effects can be added.

%mention out model assumption: point out actual cells are much
%larger, our model is minimal and coarse grained, e.g., the stress
%fiber bundles, the actual migration might be more directly
%regulated by force network

%In addition, in our model, the active particles sense force on
%indivdual bonds, in actual systems, cell sense forces on much
%larger scale, to meso-scale structure such as bundles of stress
%fibers [***bo sun PNAS***]

We note that many factors can affect the interactions between the
dynamic force network and the collective dynamics of the active
particles in our APN model. These may include the
geometry/topology and mechanical properties of the network, as
well as the number density, spatial distribution and the
contractibility (i.e., $\delta l$) of the active particles. In
this work, we mainly focus on disordered isostatic networks (i.e.,
$Z = 6$) derived from maximally random jammed packings of
congruent hard spheres \cite{ref32}. It is straightforward to
generalize this study to random network models derived from
confocal images of collagen gels \cite{ref33}. In addition, we use
simple linear elastic network models (see Methods). This allows us
to investigate the system in the elastic regime
\cite{MacKintosh05}, in which the force network is mainly
determined by the number density and spatial distribution of the
active particles, and largely independent of particle
contractibility. Our model can readily incorporate more realistic
mechanical models for the ECM, taking into account non-linear
responses of the fibers \cite{Safran12, nat_method15} and
plasticity \cite{ref29}. Moreover, in an actual cell-ECM system,
the cell migration might not be sensitive to individual stiffer
fibers, but determined by certain meso-scale stiff structures
emerged due to cell remodeling, such as bundles of high-stress
fibers. Nonetheless, we believe that the general organizational
principles of active particles on random networks obtained here
are relevant to and can provide insights on the actual cell-ECM
systems.

%%%%%%%%%%%%%%%%%%%%%$$
%first qualtatively describe the phenomena, then present a more quantiative detailed analysis

\begin{figure}[ht]
\includegraphics[width=0.48\textwidth,keepaspectratio]{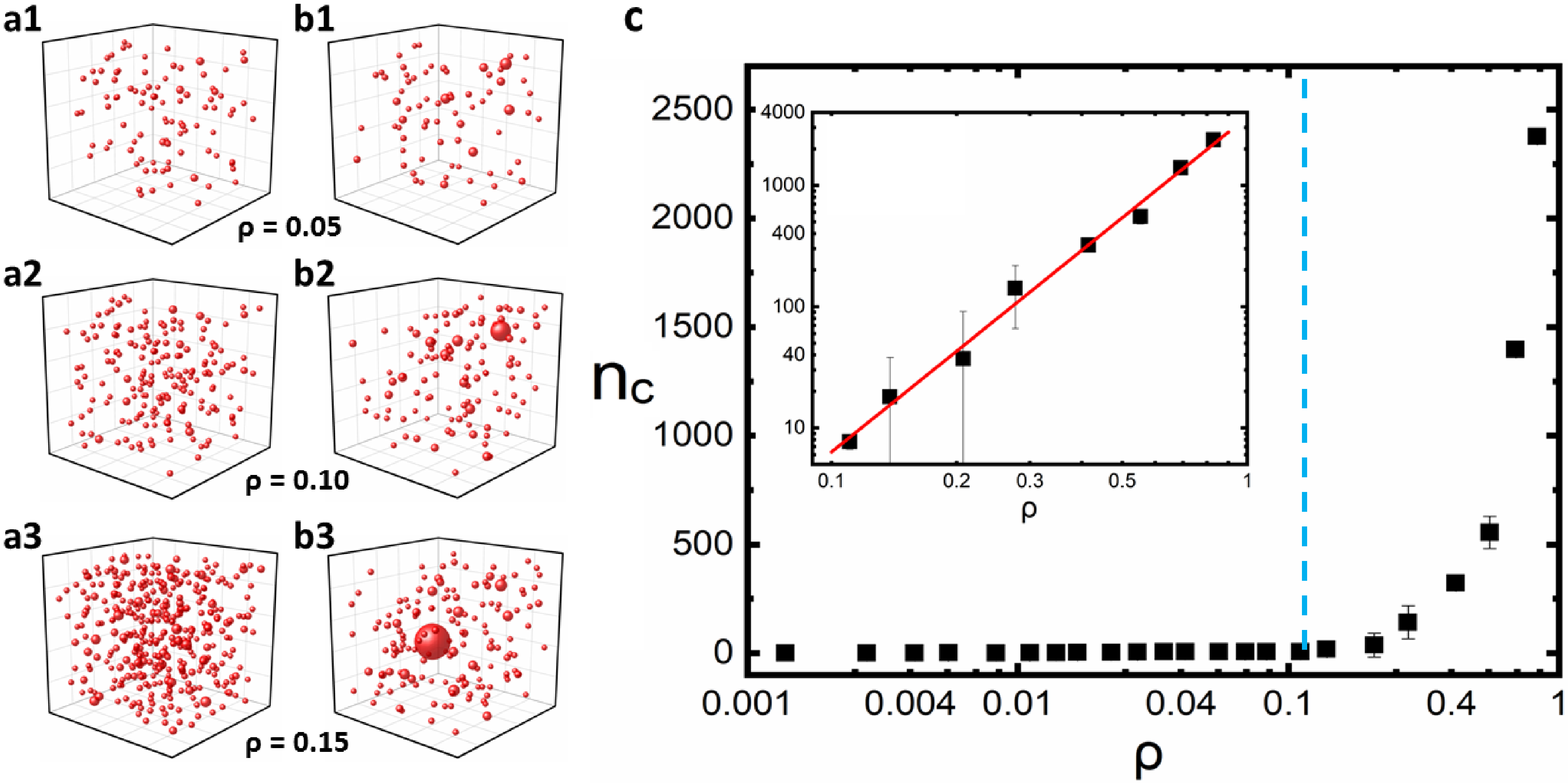}
\label{fig_2}\caption{(a) Initial distribution of clusters formed
by randomly placed active particles on the network for different
number densities $\rho$. In these plots, a cluster is represented
by a sphere for better visualization with the center coinciding
with the center of the cluster and the radius representing the
cluster size. (b) Distribution of clusters in the final state of
the APN system. As $\rho$ increases, the majority of particles
tend to aggregate into a single large cluster in the system. (c)
The maximal cluster size $n_c$ (see definition in the text) as a
function of $\rho$. A transition behavior is apparent as $\rho$
approaches $\rho_c \approx 0.114$ from below (indicated by the
dashed line). The inset shows the log-scale plot of $n_c$ for
$\rho>\rho_c$. The statistics are obtained by averaging over 20
independent simulations. }
\end{figure}

{\bf Dynamic phase transition in the APN system:} We now describe
the observed collective dynamics of the active particles on the
random networks. In our simulations, we systematically vary the
particle number density $\rho \in (0.05, 0.95)$. For each $\rho$,
the particles are initially randomly introduced in the network and
the system is allowed to evolve according to the aforementioned
APN dynamics. At low densities (i.e., $\rho <\rho_c \approx
0.114$), the particles rapidly aggregates into multiple isolated
small clusters, which are randomly distributed within the ECM (see
Fig. 2a and 2b). Here, we consider two particles belong to the
same ``cluster'' if they occupy the two nodes connected by the
same bond. As $\rho$ approaches $\rho_c$ from below, the maximal
cluster size $n_c$ (i.e., the number of particles in the largest
cluster of the system) increases dramatically (see Fig. 2c),
indicating the majority of the particles are connected to form a
single large cluster in the system.

\begin{figure}[ht]
\includegraphics[width=0.48\textwidth,keepaspectratio]{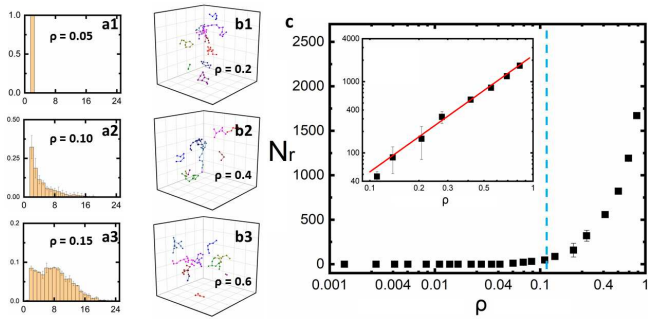}
\label{fig_3}\caption{(a) Statistics of the number of distinct
nodes $m_s$ visited by a particle for $s=24$ successive steps for
different number densities $\rho$. As $\rho$ approaches $\rho_c$
from below, a subset of highly dynamic particles emerge which are
able to visit many distinct nodes for a given number of steps and
are referred to as ``radicals''. (b) Representative trajectories
for 10 randomly selected radicals (i.e., highly dynamic particles)
for different $\rho$. (c) The number of radicals $N_r$ as a
function of $\rho$, which exhibits a clear transition at $\rho_c
\approx 0.114$ (indicated by the dashed line). This is consistent
with the transition observed in the maximal cluster size $n_c$ as
$\rho$ increases (see Fig. 2c). The inset shows the log-scale plot
of $N_r$ for $\rho>\rho_c$.}
\end{figure}

In addition, we find that the isolated small clusters associated
with $\rho<\rho_c$ are stationary, i.e., the particles in the
clusters either do not move at all or hop between two adjacent
nodes (typically at the boundary of a cluster). On the other hand,
the dominant large clusters formed for $\rho>\rho_c$ undergo
constant dynamic reorganization. To further quantify the dynamics
of the clusters, we count the number of distinct nodes $m_s$
visited by a particle during a total of $s$ successive steps. The
collected statistics for different particle densities are shown in
Fig. 3a. We note that the $m_s$ statistics shown in Fig. 3a does
not depend on $s$ and we have used $s = 24$ here.

%and collect the statistics for differen rho, show in Fig.3.

%The statistics does not depend on the steps.

%Then define the radicals, show the typical trajectory,

%Then show the number of radicals, as function of number density.

%Fig.3 shows the distribution of

It can be seen from Fig. 3a that for small $\rho$, a particle can
only visit one or two nodes, respectively indicating that the
particle does not move or can hop between two adjacent nodes. As
$\rho$ approaches $\rho_c$ from below, although the majority of
particles are localized (indicated by the peak in the $m_s$
statistics associated with small node numbers), a subset of highly
dynamic particles emerge which are able to visit many distinct
nodes for a given number of steps (indicated by the emergence of
the second peak associated with large node numbers in the $m_s$
statistics). We refer to these highly dynamic particles as
``radicals'', i.e., those do not possess a periodic hopping
pattern overall a finite number nodes. The trajectory of a small
number of randomly selected radicals are shown in Fig. 3b for
different $\rho$ values. Fig. 3c shows the number of radicals
$N_r$ as a function of $\rho$. It can be seen that $N_r$ exhibits
a clear transition as $\rho$ increases towards $\rho_c$. This is
consistent with the transition observed in the maximal cluster
size $n_c$ as $\rho$ increases (see Fig. 2c).

%Such local particle dynamics is found to occur at the boundaries
%of the well-defined clusters, which does not affect the
%identification of the clusters. For systems with $\rho > \rho_c$,
%the $\Gamma$ statistics become a continuous spectrum (see Fig.
%2c), indicating the majority of the particles are continuously
%moving from one node to another, yet still maintain a large
%dynamic cluster (see Fig. 2b).

%Provide the evidence for this, cluster statistics, dynamic
%relaxation time scale

%present the cluster statistics, show more dynamic features, using
%the average cluster size as one indicator for the transition, use
%the evolution time as another indicator.

The above analysis suggests that the system possesses a
phase-transition-like behavior, as the particle number density
$\rho$ increases, from an ``absorbing'' state in which the
particles segregate into small isolated stationary clusters, to a
``active'' state, in which the majority of particles join in a
single large dynamic cluster. This transition is also
quantitatively manifested in the maximal cluster size $n_c(\rho)$
(Fig. 2c) and radial number $N_r(\rho)$ (Fig. 3c) as $\rho$
increases towards $\rho_c \approx 0.114$. In particular, our
scaling analysis shows that as $\rho_c$ is approached from above,
$n_c \sim (\rho-\rho_c)^\alpha$, where the critical exponent
$\alpha \approx 6.6\pm0.1$. In addition, we find that $N_r \sim
(\rho-\rho_c)^\beta$, where the critical exponent $\beta \approx
1.63\pm0.04$. The numerical values of $\alpha$, $\beta$ and
$\rho_c$ are obtained by fitting the simulation data. We also note
that the $\rho_c$ value is much lower than the site percolation
threshold for the network ($\approx 0.310$) \cite{percolation}.
The absorbing-to-active transition has also been observed in a
wide spectrum of ``random-organizing'' physical systems, such
periodically driven colloids \cite{randorg01, randorg02,
randorg03}, granular materials and amorphous solids
\cite{randorg04, randorg05, randorg06}, vortices \cite{randorg07}
and skyrmion systems \cite{randorg08}.

\begin{figure}[ht]
\includegraphics[width=0.375\textwidth,keepaspectratio]{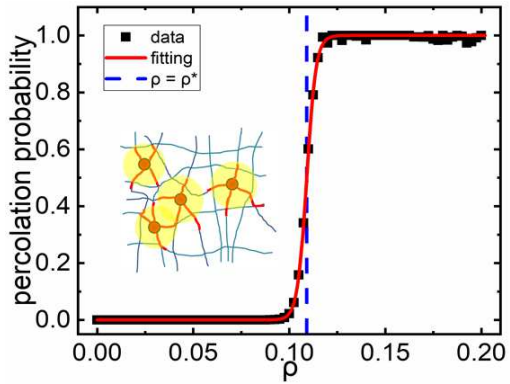}
\label{fig_4}\caption{Percolation probability analysis indicates a
percolation transition of overlapping influence spheres with
radius $R_I = 0.104L$ at $\rho^* \approx 0.109$, which agrees well
with the critical density $\rho_c \approx 0.114$ for the dynamic
phase transition in the APN system. Inset: Schematic illustration
of the concept of the influence region (yellow circles),
characterizing the range of the pulling forces (red) due to
particle (red) contraction.}
\end{figure}

%(b) Distribution of intra-cluster nearest-neighbor distance $d_n$,
%from which the influence sphere radius $R_I$ is estimated.

%show the correlation of force network and particle flux in Fig. 3
%highlighting the role of dynamic evolving force network.

{\bf Influence sphere due to active pulling forces:} We now
investigate the mechanisms for the observed transition. Once a
particle pulls the fibers, a stress gradient is built up
surrounding this particle. When another particle ``senses'' the
pulling force \cite{ref10}, it will tend to move up the stress
gradient towards the contracting particle due to local durotaxis.
This would lead to an effective mutual pulling between the
particles.

%In fact, this result is consistent with the recent observation in
%{\it in vitro} experiments on collective behaviors of invasive
%breast cancer cells.

%Then based on the force analysis (in a static fahsion, don't show
%evolution?), propose pulling force influence sphere idea: force
%decay as distance increase, force needs to be large enough to
%substaintially influence cell migration, this specify a range of
%influence; one can also get this estimates from statistics of
%isolated clusters, and plug this back into percolation theory to
%estiamte the transition.

It is reasonable to assume that the pulling forces generated by a
specific particle can only influence other particles within a
certain distance $R_I$. Due to the intrinsic network
heterogeneity, $R_I$ may vary for different particles. Here, we
take a ``mean-field'' approach and assign the same effective $R_I$
to all the particles in the system and introduce the concept of
the {\it influence sphere}, which is a spherical region with
radius $R_I$ centered at a contractile particle (see the inset of
Fig. 4).

%The idea is that at such low density, particles whose influcens
%sphere overlap can aggrege.

%Extract the pulling force influence sphere size from the data,
%point out it is difficult to predict this due to network
%heterogeneity etc. Also the statistics should be obtained from
%isolated clusters

%We also extract RI mean separation also by considering the
%dynamics of a pair of particles random placed in the network, the
%results are consistent with the cluster analysis.

The influence-sphere radius $R_I$ is estimated from the cluster
statistics of the APN systems at low $\rho$, i.e., those
containing multiple small isolated stationary clusters in the
final state. This is based on the assumption at low $\rho$, only
particles which are within the influence region of one another
would eventually aggregate. In particular, we first identify the
particles within the same cluster in the final state of the
system. Then the system is ``re-winded'' to the initial state, and
the intra-cluster nearest-neighbor distances $d_n$ are computed
for all clusters. We then use the mean nearest-neighbor distance
$\bar d_n$ to estimate $R_I \approx 0.104L$ (where $L$ is the
linear system size), which is roughly twice of the average fiber
length (see SI for fiber length distributions).

%(i.e., the shortest distance between a specific particle and any
%other particles in the same cluster)

%Fig. 4b shows the $d_n$ statistics for systems with different
%$\rho$ values, which are statistically similar to one another.

%Then discuss the geometrical percolation results of the influence
%sphere. Show the connections.

%(i.e., two overlapping spheres are considered to belong to the
%same cluster)

%and determine the maximal cluster size $\ell_c$ (i.e., the linear
%dimension of the largest cluster formed by the overlapping virtual
%influence spheres along the three orthogonal directions, in the
%unit of the edge length of the simulation box),

{\bf Mean field theory: Percolation of influence sphere:} We now
investigate the percolation of the influence spheres as $\rho$
increases. For a given $\rho$, we randomly place particles on the
ECM network. Instead of allowing the particles to move according
to the APN dynamics, we place a virtual sphere with radius $R_I =
0.104L$ at each particle, representing the influence spheres. We
subsequently identify the clusters formed by the influence spheres
, based on which the percolation of the system can be determined.

Fig. 4 shows the percolation probability analysis for the system
\cite{ref34} (see SI for details), from which a percolation
transition and the associated critical density (i.e., percolation
threshold) $\rho^* \approx 0.109$ can be clearly identified.
Interestingly, the percolation transition of the influence spheres
coincides with the dynamic transition of the active particles at
$\rho_c \approx 0.114$. This suggests that the dynamic transition
of the active particles from the ``absorbing'' state to the
``active'' state is underlaid by and can be understood as the
percolation transition of the influence spheres.

% as a function of
%$\rho$, which clearly indicates a percolation transition as $\rho$
%increases. The percolation threshold $\rho^* \approx XXXX$ is
%obtained via percolation probability analysis \cite{ref34} (see SI
%for details).

%Finally, talk about other networks, also mention that the
%contractiability of the cell does not significantly affect the
%results, put the results in the SI. Point out the physics is
%robust, phenomena may vary a little. Point out the implementation
%in real system, mention the difference, e.g., we don't consider
%cell-cell adhesion and ECM degradation.

We also investigate the collective dynamics of active particles on
other network models, including the networks derived from the
diamond lattice, as well as random networks reconstructed based on
confocal images (see SI for details). We find that although the
critical transition density $\rho_c$ depends on the geometry and
topology of the networks, the dynamic transition and the
aggregation behaviors of the active particles occur in all of the
studied model networks. This suggests the feedback loop between
the evolving force network and particle dynamics provides a robust
mechanism for regulating collective migratory behaviors. In future
work, we will also explore the effects of fiber alignment and
external mechanical cues.

%This suggests that

%Mention the key point: The migration of particles constantly
%re-shape the force network, which also depends on the original
%network topology and geometry. The particles then move in response
%to the force network via local durotaxis.

Finally, we emphasize again that this minimal model does not take
into account crucial mechanisms in actual cell migration such as
ECM remodeling (e.g., orientation, bundling and degradation) and
cell-cell adhesion. Interestingly, our studies indicate that, at
least for the APN systems, the local durotaxis for the active
particles is sufficient to induce and stabilize aggregations, even
without adhesion. Nonetheless, we expect that the insights on the
collective behaviors of active particles regulated by the
dynamically evolving force network obtained here are helpful in
understanding the collective dynamics emerged in actual
multi-cellular-ECM systems, as well as in other active-particle
systems dominated by environment-mediated particle-particle
interactions.

\begin{acknowledgments}
%\bigskip
H. N., Y. Z., Y. J. thank Arizona State University for the
generous start-up funds and the University Graduate Fellowship.
W.X. was supported by the National Natural Science Foundation of
China (Grant No. 11772120). C. E and B. S. thank the support from
the Scialog Program sponsored jointly by Research Corporation for
Science Advancement and the Gordon and Betty Moore Foundation. B.
S. is partially supported by the Medical Research Foundation of
Oregon and SciRIS-II award from Oregon State University and by the
National Science Foundation Grant PHY-1400968.

%\noindent{\bf Data Availability} All data and computer codes are
%available from the authors upon reasonable request.

%\noindent{\bf Author Contributions} Y. J., B. S., and W. X.
%designed and oversaw the research. H. N., Y. Z., S. C., C. E., J.
%T. performed the research and collected data. All authors analyzed
%data and wrote the manuscript.

\end{acknowledgments}

%\bibliography{network}

\end{document}